\documentclass[a4paper,11pt]{article}
\usepackage{pos}

\usepackage{slashed} 
\usepackage{hyperref}

\def\bom#1{{\mbox{\boldmath $#1$}}}

\def\l{\left}

\def\eqv{\equiv}

\def\ep{\epsilon}

\def\wt{\widetilde}
\def\fr{\frac}

\def\beqn{\begin{eqnarray}} \def\eeqn{\end{eqnarray}}
\def\beq{\begin{equation}} \def\eeq{\end{equation}}

\newcommand{\la}{\langle}
\newcommand{\ra}{\rangle}
\def\sp{{\bom {Sp}}}
\def\ket#1{|{#1}\ra}

\def\wp{\widetilde P}

\title{
\vspace*{-2.5cm}
\begin{minipage}{\textwidth}
{\normalfont\small IFIC/23-47
\hspace{\fill} October 2023
}\\
\end{minipage}\\[60pt]
  Towards higher-order collinear splittings with massive partons}

\ShortTitle{Towards higher-order collinear splittings with massive partons}

\author*[a,b]{German F. R. Sborlini}
\author[c]{Prasanna K. Dhani}
\author[c]{Germ\'an Rodrigo}

\affiliation[a]{Departamento de F\'isica Fundamental e IUFFyM, Universidad de Salamanca, 37008 Salamanca, Spain.}
\affiliation[b]{Escuela de Ciencias, Ingenier\'ia y Diseño, Universidad Europea de Valencia, Paseo de la Alameda 7, 46010 Valencia, Spain.}
\affiliation[c]{Instituto de F\'{\i}sica Corpuscular, Universitat de Val\`{e}ncia -- 
Consejo Superior de Investigaciones Cient\'{\i}ficas, Parc Cient\'{\i}fic, 46980 Paterna, Valencia, Spain.}

\emailAdd{german.sborlini@usal.es}
\emailAdd{dhani@ific.uv.es}
\emailAdd{german.rodrigo@csic.es}

\abstract{The singularities associated with QCD factorization in the collinear limit are key ingredients for high-precision theoretical predictions in particle physics. They govern the collinear behaviour of scattering amplitudes, as well as the perturbative energy evolution of parton densities (PDFs) and fragmentation functions (FFs). In this talk, we present the computation of multiple collinear and higher-order QCD splittings with massive partons. Our results are highly-relevant for the consistent introduction of mass effects in the subtraction formalism and PDF/FF evolution.}

\FullConference{The European Physical Society Conference on High Energy Physics (EPS-HEP2023)\\
 21-25 August 2023\\
Hamburg, Germany\\}

\begin{document}
\maketitle

\section{Introduction}
\label{sec:Introduction}
Considerable progress has been made in comprehending the factorization properties of hard-scattering matrix elements involving massless partons. However, the scenario changes when dealing with massive partons, demanding more dedicated endeavors, particularly in light of current (and future) experiments in particle physics. The high-energy colliders, in their quest to understand Standard Model (SM) and physics beyond it, have shown a pronounced interest in the production of heavy quarks. In order to enhance the accuracy of theoretical predictions, it is imperative to account for higher-order contributions beyond the leading order (LO) within perturbation theory. Equally critical is the need to effectively manage explicit infrared (IR) poles and potentially substantial logarithmic contributions resulting from scale hierarchies \cite{Catani:2022sgr}. The application of QCD factorization with massive particles emerges as a central element in the pursuit of these objectives.

It is well-known that radiation from massive partons is strongly suppressed in the collinear limit by virtue of a phenomenon known as the dead-cone effect. Additionally, explicit collinear poles in perturbative calculations are prevented by the presence of mass. However, they can still result in IR finite contributions that scale as $\ln^p(Q^2/m^2)$, where $Q$ represents the typical energy scale of the hard-scattering process and $m$ signifies the mass of the heavy parton. In situations where $Q^2 \ll m^2$, these substantial logarithmic contributions have the potential to detrimentally affect both the numerical convergence and the reliability of theoretical predictions for a wide range of observables.

The encouraging aspect is that these contributions can be computed in a manner that is independent of the specific process \cite{Catani:2022sgr} and can be resummed to all orders within perturbation theory, as they are linked to the singular behavior of matrix elements as the mass approaches zero. This singular behavior is governed by QCD factorization in the quasi-collinear limit, much like how soft and collinear factorization control IR divergences. At ${\cal O}(\alpha_S)$, double-parton quasi-collinear limits for various tree-level splitting processes were calculated in earlier works \cite{Keller:1998tf,Catani:2000ef}.

In this talk, we study simultaneous collinear limit of three partons in sight of the importance for accurately predicting dominant mass effects in phenomenological analysis. We calculate the full set of splitting amplitudes and kernels involved in QCD and QED processes, which were presented for the first time in Ref. \cite{Dhani:2023uxu}. At the squared amplitude level, we also entirely consider azimuthal correlations, which are distinctive of the gluon splittings and are indispensable in devising general procedures for the precise computation of jet cross sections involving massive particles at next-to-next-to-leading order (NNLO).

\section{Collinear and quasi-collinear limits}
\label{sec:CollQuasicoll}
Let us briefly describe the multiple collinear limit of scattering amplitudes and review the concept of \emph{splitting amplitudes} \cite{Catani:1999ss,Catani:2011st,Sborlini:2013jba,Catani:2022sgr}. We consider a scattering process at tree level with $N$-external particles and with a subset $C=\{1,\ldots,m\}$ composed by particles emitted in the same direction; i.e. $m$~particles become simultaneously collinear. We denote by $\ket{{\cal M}^{(0)}\left(p_1, \ldots, p_m, \ldots,  p_N\right)}$ the corresponding scattering amplitude, with $p_i^{\mu}$ being the four-momentum associated with particle~$i$ and~$p_{1\ldots m} = p_1 + \ldots + p_m$. We will denote with $a_i$ the flavour of particle $i$, although we will avoid indicating this explicitly unless it is strictly required. To continue the description of the collinear emission, it is useful to introduce the ancillary light-like vector $n$ and define
\beq
\wp^{\mu} = p_{1\ldots m}^{\mu} - \frac{p^2_{1 \ldots m}}{2 \; n \cdot p_{1\ldots m}} \, n^{\mu}~, 
\label{eq:Ptilde}
\eeq
which corresponds to the collinear on-shell direction, where $s_{1\ldots m}\equiv p^2_{1 \ldots m}= \left(p_1 + \ldots + p_{m} \right)^2$ and $\wp^2=0$ by construction. In this way, the vector $n$ indicates how we are approaching to the strict collinear limit. The next step consists in adopting the Sudakov parametrization for the collinear momenta, i.e.
\beq
p_i^{\mu} = z_i \wp^{\mu} + k_{\perp i}^{\mu} - \fr{k_{\perp i}^2}{2 \, z_i \, n \cdot \wp}\, n^{\mu}~, \qquad i \in C~,
\label{eq:SudakovMassless}
\eeq
where $z_i$ represents the momentum fraction carried by the particle $i$ in the collinear direction and $k_{\perp i}$ the projection of $p_i$ in the space transverse to $n$ and $\wp$. Then, we have the useful identities $n\cdot k_{\perp i} = 0$ and $\wp \cdot k_{\perp i} = 0$, as well as
\beq
\sum_i z_i =1~, \qquad  \sum_i k_{\perp i}^\mu = 0~.
\eeq
With this parametrization, we define the collinear limit as $p^2_{1\ldots m}\to 0$, or alternatively by performing the rescalling $\lambda \, k_{\perp i}$ for $i \in C$ and taking the limit $\lambda \to 0$. Due to strict collinear factorization properties \cite{Collins:1989gx,Catani:2011st}, the tree-level amplitude factorizes according to
\beq
\ket{{\cal M}^{(0)}\left(p_1, \ldots, p_N\right)} \approx 
\sp^{(0)}_{a_1 \dots a_m}(p_1 , \ldots, p_m; \wp) \, 
\ket{{\cal M}^{(0)}(\wp, p_{m+1}, \ldots, p_N)} \, , 
\label{eq:FactorizationMassless}
\eeq
which is valid for the most singular terms in the limit $s_{1 \ldots m} \to 0$. The splitting amplitude $\sp^{(0)}_{a_1 \dots a_m}(p_1 , \ldots, p_m; \wp)$ is a universal factor that fully embodies the singular behaviour of any scattering amplitude when the parent particle $a$ with momenta $\wp$ undergoes a collinear splitting into the particles $\{a_i\}_{i\in C}$, and only depends on information carried by these collinear particles.

Now, we proceed to characterize splitting processes in which massive particles are involved. In first place, it is straightforward to prove that collinear singularities are absent, since the mass acts as a regulator. Still large-logarithmic corrections might spoil the numerical convergence. At this point, let us consider the same matrix element introduced in the beginning of this section, and lift off the massless restriction. Namely, we consider that each particle $a_i$ within the subset $C$ has an arbitrary mass $m_i$. The mass of the parent particle $a$, which undergoes the splitting, is $m_{1\ldots m}$. Since this parent particle is generally off shell because $p_{1\dots m}^2 \neq m_{1\ldots m}^2$, we define the on-shell vector
\begin{align}
    \wt{P}^{\mu} = p_{1\dots m}^{\mu}-\frac{p_{1\ldots m}^2-m_{1\ldots m}^2}{2\, n \cdot p_{1\ldots m}}n^{\mu} \, ,
    \label{eq:Ptildegeneral}
\end{align}
which fulfils $\wt{P}^2 = m_{1\ldots m}^2$. Then, we modify the Sudakov parametrization given in Eq. (\ref{eq:SudakovMassless}) according to
\begin{align}
    p_i^{\mu} = x_i \wt{P}^{\mu} + k_{\perp i}^{\mu} - \fr{k_{\perp i}^2+x_i^2m_{1\ldots m}^2-m_i^2}{x_i}\frac{n^{\mu}}{2\,n\cdot \wt{P}}\ , \qquad i \in C~,
\end{align}
where we defined 
\beq
z_i = \fr{x_i}{\sum_{j=1}^{m} x_j} , \qquad \tilde{k}_i^{\mu} = k_{\perp i}^{\mu}-z_i\sum_{j=1}^{m} k_{\perp j}^{\mu}~.
\eeq
Notice that we recover the usual notion of momentum fraction if we set $\sum_i x_i = 1$. Also, we can distinguish
\beqn
    s_{ij} &\eqv& (p_i+p_j)^2 = -z_i z_j\left( \fr{\tilde{k}_j}{z_j}-\frac{\tilde{k}_i}{z_i}\right)^2 + (z_i+z_j)\left(\frac{m_i^2}{z_i}+\frac{m_j^2}{z_j} \right)~,
    \label{eq:pijmassive}
\eeqn
and
\beqn    
    \tilde{s}_{ij}&\eqv& 2p_i\cdot p_j = z_i z_j\left[-\left( \fr{\tilde{k}_j}{z_j}-\frac{\tilde{k}_i}{z_i}\right)^2 + \fr{m_i^2}{z_i^2} + \fr{m_j^2}{z_j^2} \right]~,
    \label{eq:sijmassive}
\eeqn
which are equivalent only in the massless case. It turns out that, by virtue of Eq. (\ref{eq:pijmassive}), the invariant mass of the splitting system depends on $\{k_{\perp i}^2,m_i^2\}$ and $m_{1\ldots m}^2$. So, this allow us to formally define the \emph{quasi-collinear} kinematical region by uniformly rescalling these variables, i.e.
\beq
m_i \to \lambda \, m_i~, \qquad k_{\perp i} \to \lambda \, k_{\perp i}~, \qquad m_{1\ldots m} \to \lambda \, m_{1\ldots m}~,
\eeq
and taking the limit $\lambda \to 0$. Keeping the most singular terms in $\lambda$, we recover the factorization formula in Eq. (\ref{eq:FactorizationMassless}). Even if the structure is formally the same, there are some relevant differences. For instance, the massive splittings have an implicit (and often explicit) dependence on the mass, embodied in the scalar products involving $p_i$ for massive partons.

\section{Massive triple collinear splittings}
\label{sec:MassiveSplittings}
In this section, we present selected unpolarized splitting kernels in the triple collinear limit. We limit ourselves to report the expressions shown in Ref. \cite{Dhani:2023uxu}, following the notation introduced there. For the quark-initiated process with two different quark flavours, we obtain
\beqn
\nonumber \la\hat{P}^{(0)}_{\bar{Q}_1^\prime Q_2^\prime Q_3}\ra  &=& C_F T_R\Bigg\{\fr{\tilde{s}_{12}\tilde{s}_{123}}{2s_{12}^2} \left[-\fr{t_{12,3}^2}{\tilde{s}_{12}\tilde{s}_{123}}+\fr{4z_3+(z_1-z_2)^2}{1-z_3}+(1-2\ep)\left(z_1+z_2-\fr{\tilde{s}_{12}}{\tilde{s}_{123}}\right)\right]
\\ \nonumber &+& \fr{2m_{Q^\prime}^2}{s_{12}^2}\bigg[\fr{z_3 \tilde{s}_{123} }{(1-z_3)^2}(1+2z_3-3z_3^2+4z_1z_2)-\fr{\tilde{s}_{23}}{1-z_3}(2-3z_1-5z_2+z_1^2+z_2^2)  
\\ \nonumber &-& \fr{\tilde{s}_{13}}{1-z_3}(2-5z_1-3z_2+z_1^2+z_2^2)-\ep\Big(\tilde{s}_{123}(1-z_3)-\tilde{s}_{12}(1+z_3)\Big)\bigg]-2\fr{m_{Q}^2\tilde{s}_{12}}{s_{12}^2}
\\ &+& \fr{4m_{Q^\prime}^4}{s_{12}^2}z_3\left[\ep+\fr{2z_1z_2}{(1-z_3)^2}+\fr{2z_3}{1-z_3}\right]-4\fr{m_{Q}^2m_{Q^\prime}^2}{s_{12}^2}\Bigg\}~,
\label{eq:Q2QpbQpQ}
\eeqn
where
\beq
t_{ij,k} = 2 \frac{z_i \tilde{s}_{jk}-z_j \tilde{s}_{ik}}{z_i+z_j} + \frac{z_i -z_j}{z_i+z_j} \tilde{s}_{ij} \, ,
\label{eq:Variables}
\eeq
is a kinematical variable, first introduced in Ref. \cite{Catani:1999ss} since it leads to a very compact expression.

Also, we present the unpolarized gluon-initiated splitting kernel, given by
\begin{align}
    \la\hat{P}^{(0)}_{g_1Q_2\bar{Q}_3}\ra = C_FT_R \, \la\hat{P}^{\rm{(0,ab)}}_{g_1Q_2\bar{Q}_3}\ra + C_AT_R\, \la\hat{P}^{\rm{(0,nab)}}_{g_1Q_2\bar{Q}_3}\ra~,
\end{align}
with
\beqn
\nonumber \la\hat{P}^{\rm{(0,ab)}}_{g_1Q_2\bar{Q}_3}\ra &=& \Bigg\{\left[ \frac{s_{123}^2 \left(z_1^2 (1-\epsilon )+2 (1-z_2) z_3-\epsilon \right)}{\tilde{s}_{12} \tilde{s}_{13}(1-\epsilon)} + \frac{2 s_{123} ((z_1+1) \epsilon +z_2-1)}{\tilde{s}_{12}(1-\epsilon)}-\epsilon\right.
\\ \nonumber &+& \left. \frac{\tilde{s}_{13} (1-\epsilon )}{\tilde{s}_{12}} \right] + \left[ \frac{m_Q^2}{\tilde{s}_{12}} \,\left(\frac{2 s_{123} (2 (1-z_3) z_3+\epsilon -1)}{\tilde{s}_{12}}+\frac{2 s_{123} (z_1+2 z_2 z_3+\epsilon )}{\tilde{s}_{13}}-4 \right) \right.
\\ &-& \left.  \,\frac{4 \, m_Q^4}{\tilde{s}_{13}} \left(\frac{1}{\tilde{s}_{12}}+\frac{1}{\tilde{s}_{13}}\right) \right]\frac{1}{1-\epsilon}\Bigg\} \,   + (2\leftrightarrow 3) \, ,
\eeqn
and
\beqn
\label{eq:g2gQQpNab}
\nonumber \la\hat{P}^{\rm{(0,nab)}}_{g_1Q_2\bar{Q}_3}\ra &=&  \Bigg\{\left[  \frac{s_{123}^2 z_3 }{2 s_{23} \tilde{s}_{13}}\left(\frac{(1-z_1)^3-z_1^3}{z_1 (1-z_1)}-\frac{2 z_3 (-2 z_1 z_2-z_3+1)}{z_1 (1-z_1) (1-\epsilon )}\right) - \frac{s_{123}^2 }{2 \tilde{s}_{12} \tilde{s}_{13}}\left(z_1^2-\frac{z_1+2 z_2 z_3}{1-\epsilon }+1\right) \right.
\\ \nonumber &-& \left. \frac{(\tilde{t}_{23,1})^2}{4 s_{23}^2}+\frac{\epsilon }{2}-\frac{1}{4}+ \frac{s_{123} }{2 s_{23}}\left(\frac{z_1^3+1}{z_1 (1-z_1)}+\frac{z_1 (z_3-z_2)^2-2 (z_1+1) z_2 z_3}{z_1 (1-z_1) (1-\epsilon )}\right) \right.   
\\ \nonumber &+& \left.  \frac{s_{123} (1-z_2) }{2 \tilde{s}_{13}}\left(-\frac{2 (1-z_2) z_2}{z_1 (1-z_1) (1-\epsilon )}+\frac{1}{z_1 (1-z_1)}+1\right) \right] 
\\ \nonumber &+& \left[ \frac{m_Q^2}{s^2_{23}} \,\left(\frac{2 s_{123}^2 (1-z_2)}{\tilde{s}_{13} (1-z_1) z_1}-\frac{s_{123}^3 (z_1+2 z_2 z_3+ \epsilon )}{\tilde{s}_{12} \tilde{s}_{13}}-\frac{2 s_{123}^2 z_1^2 (1-2 z_2)}{\tilde{s}_{13} (1-z_1)} \right. \right.
\\ \nonumber &+& \left. \left. \frac{2 s_{123}^2(4 (1-z_2) z_2+z_2+2 \epsilon -2)}{\tilde{s}_{13}}-\frac{2 s_{123} \tilde{s}_{12} z_2}{\tilde{s}_{13} (1-z_1)}-\frac{2 s_{123} \tilde{s}_{12} z_3}{\tilde{s}_{13} z_1} \right. \right.
\\ \nonumber &-& \left. \left. s_{123} (z_1 (1-4 z_2)+4 (1-z_2) z_2+2 \epsilon +3)+\frac{s_{123}}{z_1} + 4 \tilde{s}_{12} \right. \right.
\\ &-& \left. \left.  \frac{2 s_{123} \tilde{s}_{12} \left(z_2-2 z_2^2+\epsilon \right)}{\tilde{s}_{13}} \right)  \, +   \,\frac{2 \, m_Q^4}{\tilde{s}_{12}\tilde{s}_{13}} \right] \frac{1}{1-\epsilon}\Bigg\} \,   + (2\leftrightarrow 3) \, .
\eeqn
In order to check the validity of our calculations, we consider the massless limit, recovering the well-known expressions available in Ref. \cite{Catani:1999ss}.

\section{Conclusions and outlook}
\label{sec:conclusions}
In this talk, we have discussed the first calculation of triple-collinear splitting amplitudes and (un)polarized splitting kernels with massive partons at tree-level in QCD \cite{Dhani:2023uxu}. We present selected results for the unpolarized kernels, corresponding to the splitting processes $Q \to Q'_1 \bar{Q}'_2 Q_3$ and $g \to g_1 Q_2 \bar{Q}_3$.

The study of the quasi-collinear limit and, in particular, the calculation of the associated multiple-collinear splittings with massive partons plays an important role in the consistent inclusion of mass effects in high-precision collider physics. These results could be used for improving the currently available subtraction frameworks, since the massive splittings are crucial for building counter-terms. Besides, the full set of triple-collinear splitting kernels with massive partons could allow to consistently include mass effects in a new parton shower generators. Finally, we could use the results provided in Ref. \cite{Dhani:2023uxu} to calculate higher-order corrections to Altarelli-Parisi kernels including mass effects, which might allow to determine the evolution of PDFs/FFs at the highest-precision taking consistently into account the masses of the heavy quarks.

\textbf{Note}: After our paper was released on arXiv (and more than a month after our results were presented at EPS-HEP 2023), an independent calculation of the quark-initiated massive triple-collinear splitting kernels was presented in Ref. \cite{Craft:2023aew} and it is in agreement with our expressions. Notice also that Ref. \cite{Dhani:2023uxu} includes full results both for splitting amplitudes and splitting kernels.

  \subsection*{Acknowledgments}
This work is supported by the Spanish Government (Agencia Estatal de Investigaci\'on MCIN /AEI/10.13039/501100011033) Grants No. PID2020-114473GB-I00, PID2022-141910NB-I00 and Generalitat Valenciana Grant No. PROMETEO/2021/071. 
The work of PKD is supported by European Commission MSCA Action COLLINEAR-FRACTURE, Grant Agreement No. 101108573.
The work of GS is partially supported by H2020-MSCA-COFUND USAL4EXCELLENCE-PROOPI-391 project under Grant Agreement No 101034371.


\begin{thebibliography}{1}

\bibitem{Catani:2022sgr}
S.~Catani and P.~K. Dhani, \emph{{Collinear functions for QCD resummations}},
  \href{http://dx.doi.org/10.1007/JHEP03(2023)200}{\emph{JHEP} {\bf 03} (2023)
  200}, [\href{http://arxiv.org/abs/2208.05840}{{\tt 2208.05840}}].

\bibitem{Keller:1998tf}
S.~Keller and E.~Laenen, \emph{{Next-to-leading order cross-sections for tagged
  reactions}}, \href{http://dx.doi.org/10.1103/PhysRevD.59.114004}{\emph{Phys.
  Rev. D} {\bf 59} (1999) 114004},
  [\href{http://arxiv.org/abs/hep-ph/9812415}{{\tt hep-ph/9812415}}].

\bibitem{Catani:2000ef}
S.~Catani, S.~Dittmaier and Z.~Trocsanyi, \emph{{One loop singular behavior of
  QCD and SUSY QCD amplitudes with massive partons}},
  \href{http://dx.doi.org/10.1016/S0370-2693(01)00065-X}{\emph{Phys. Lett. B}
  {\bf 500} (2001) 149--160}, [\href{http://arxiv.org/abs/hep-ph/0011222}{{\tt
  hep-ph/0011222}}].

\bibitem{Dhani:2023uxu}
P.~K. Dhani, G.~Rodrigo and G.~F.~R. Sborlini, \emph{{Triple-collinear
  splittings with massive particles}},
  \href{http://arxiv.org/abs/2310.05803}{{\tt 2310.05803}}.

\bibitem{Catani:1999ss}
S.~Catani and M.~Grazzini, \emph{{Infrared factorization of tree level QCD
  amplitudes at the next-to-next-to-leading order and beyond}},
  \href{http://dx.doi.org/10.1016/S0550-3213(99)00778-6}{\emph{Nucl. Phys. B}
  {\bf 570} (2000) 287--325}, [\href{http://arxiv.org/abs/hep-ph/9908523}{{\tt
  hep-ph/9908523}}].

\bibitem{Catani:2011st}
S.~Catani, D.~de~Florian and G.~Rodrigo, \emph{{Space-like (versus time-like)
  collinear limits in QCD: Is factorization violated?}},
  \href{http://dx.doi.org/10.1007/JHEP07(2012)026}{\emph{JHEP} {\bf 07} (2012)
  026}, [\href{http://arxiv.org/abs/1112.4405}{{\tt 1112.4405}}].

\bibitem{Sborlini:2013jba}
G.~F.~R. Sborlini, D.~de~Florian and G.~Rodrigo, \emph{{Double collinear
  splitting amplitudes at next-to-leading order}},
  \href{http://dx.doi.org/10.1007/JHEP01(2014)018}{\emph{JHEP} {\bf 01} (2014)
  018}, [\href{http://arxiv.org/abs/1310.6841}{{\tt 1310.6841}}].

\bibitem{Collins:1989gx}
J.~C. Collins, D.~E. Soper and G.~F. Sterman, \emph{{Factorization of Hard
  Processes in QCD}},
  \href{http://dx.doi.org/10.1142/9789814503266_0001}{\emph{Adv. Ser. Direct.
  High Energy Phys.} {\bf 5} (1989) 1--91},
  [\href{http://arxiv.org/abs/hep-ph/0409313}{{\tt hep-ph/0409313}}].

\bibitem{Craft:2023aew}
E.~Craft, M.~Gonzalez, K.~Lee, B.~Mecaj and I.~Moult, \emph{{The 1
  $\rightarrow$ 3 Massive Splitting Functions from QCD Factorization and
  SCET}},  \href{http://arxiv.org/abs/2310.06736}{{\tt 2310.06736}}.

\end{thebibliography}

\providecommand{\href}[2]{#2}\begingroup\raggedright\endgroup

\end{document}